# An Optimization-Accelerated Electromagnetic Time Reversal-based Fault Location Method for Power Lines with Branches

Guanbo Wang, Chijie Zhuang, Rong Zeng

*Abstract*—It is very important to locate the short-circuit fault in a power system quickly and accurately. Electromagnetic time reversal (EMTR) has drawn increasing attention because of its clear physical background and excellent performance. This paper studies the EMTR method for locating the short-circuit fault of transmission and distribution lines with or without branches, and introduces a simulated annealing algorithm to accelerate the calculation of an EMTR fault location. This algorithm is different from the traditional exhaustive method in that it solves the corresponding optimization problem, thus improving the location speed by up to an order of magnitude. With the help of graph theory, a method is proposed that automatically splits a complex line topology with branches into several one-dimensional lines. The problem of short-circuit fault location in the branching lines is then transformed into several one-dimensional optimization problems, which are then solved by the optimization algorithm. This solves the problem of realizing rapid location in a power network with branches. Numerical experiments are carried out in a distribution network model to demonstrate the effectiveness of the method. Results under different conditions show the method works reliably and efficiently.

*Keywords*—Electromagnetic time reversal, fault location, optimization algorithm, branched transmission line, graph theory

## I. Introduction

Electric energy plays a vital role in the normal operation of society and people's lives. It is very important to locate the fault of a short circuit in a power system quickly and accurately for the continuous and reliable supply of electric energy so that the fault can be eliminated in time, thereby avoiding further disruption and loss.

Existing fault location methods can be divided into at least three categories: impedance-based [3][4], traveling wave-based [5]-[7], and artificial intelligence-based [8]-[10]. The impedance method is widely used because of its simplicity and low cost. However, the method is affected by the fault impedance and the network structure. The traveling-wave method has better accuracy but requires high-frequency signal acquisition equipment, which results in more cost. The artificial intelligence method needs a lot of historical data to train the model, and it is still in the developmental stage with limited practical application. In recent years, electromagnetic time reversal (EMTR) has been proposed to overcome the difficulties and has been extensively researched.

EMTR has a clear physical background and excellent anti-noise performance compared with the impedance and traveling-wave methods. The time-reversal method was proposed by Bogert of Bell Laboratories and applied to the field of electronics [2], and later to more fields such as bio-medicine [16], imaging [17]-[19], and seismic monitoring [20].

The EMTR method was used to locate faults in transmission lines around 2012 [11]. Razzaghi et al. have since conducted a series of studies on the practicability of EMTR fault location. A fault location test was carried out on a complex T-network of components with different wave impedances, and the locating accuracy was analyzed [12]. The test confirmed that accurate location of short-circuit faults in a complex T-network only requires a single-ended signal measurement. Because the telegraphic equation of a transmission line, which is the physical basis of EMTR fault location, does not satisfy time-reversal invariance when the transmission line loss caused by line resistance is considered, the line loss was considered to test its effect on the locating accuracy [13]-[15]. The test showed that the line loss has little influence on the accuracy of fault location in practice. In addition, experiments of EMTR fault location have been carried out in recent years. Wang et al. performed fault location experiments in medium-voltage distribution networks in China and Switzerland [16][17]. The results showed that the fault locations were correctly identified via EMTR with high accuracy, e.g., within 10 m. Overall, promising results for EMTR fault location have been obtained both in numerical simulation and real experiments.

The basic principle of EMTR fault location is that locating a fault is equivalent to acquiring the maximum short-circuit current energy. In most existing studies, it has been necessary to set up a series of guess short-circuit branches at all possible positions along the line and calculate the respective short-circuit current energies. All EMTR methods thus far have used exhaustive searching (namely, brute-force searching), resulting in a large computational load.

In this paper, we propose a global optimization algorithm to accelerate fault location via EMTR, with special focus on power lines with branches because it is difficult to traverse all positions in a network using a single variable.

The rest of the paper is organized as follows: the existing EMTR fault location methods are briefly reviewed in Section II; then, Section III introduces a one-dimensional optimization algorithm instead of the exhaustive process for non-branched transmission lines to greatly improve the calculation speed of classical EMTR fault location; in Section IV, we show a way to decompose a complex network into one-dimensional paths, thereby transforming the whole problem into a superposition of problems finding the global maxima of the unary functions, which realizes EMTR fault location for transmission lines with branches. An example of faults in a distribution network is given in Section V to

Wang, Zhuang, and Zeng are with Department of Electrical Engineering, Tsinghua University, Beijing 100084, China. Zhuang is also with International Joint Laboratory on Low Carbon Clean Energy Innovation, Tsinghua University, Beijing100084, China.
(corresponding author: Chijie Zhuang. e-mail: chijie@tsinghua.edu.cn).

demonstrate the effectiveness and efficiency of the method. Finally, some conclusions are drawn in Section VI.

## II. BASIC PRINCIPLE OF THE EXISTING EMTR FAULT LOCATION METHOD

The basic principle of the classical EMTR fault location method is given in a series of papers [11]-[15]. A short-circuit fault occurs at $x = x_f$ in Fig. 1(a) and is represented by a voltage source $U_f$.

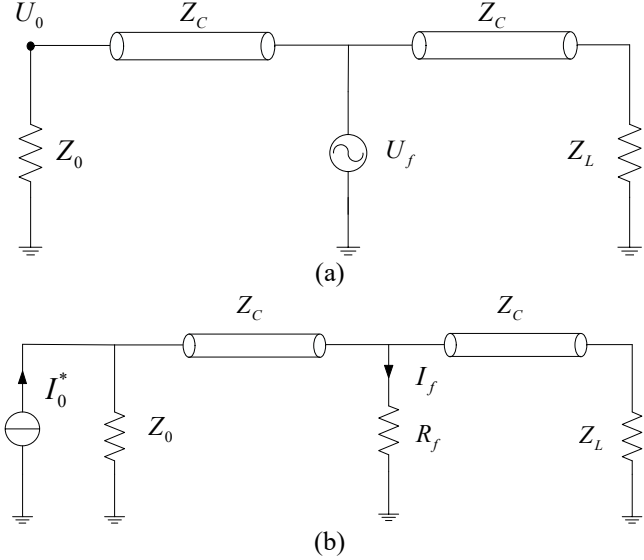

(a)

(b)

Fig. 1. Representation of a short-circuit fault along a transmission line and the time-reversal process.

A transient electromagnetic signal generated by the fault is collected at one end of the line (in this case $U_0$ is measured). The expression for $U_0$ in the frequency domain is

$$U_0(\omega) = \frac{(1+\rho_0)e^{-\gamma x_f}}{1+\rho_0 e^{-2\gamma x_f}} U_f(\omega), \quad (1)$$

where $\gamma$ is the line propagation constant, $\gamma = j\beta = j\frac{\omega}{c}$ (c is the propagation speed) without considering loss, and $\rho_0$ is the reflection coefficient given by

$$\rho_0 = \frac{Z_0 - Z_C}{Z_0 + Z_C}. \quad (2)$$

The signal is re-injected into the original system using the Norton equivalent from the same point after the time-reversal operation as shown in Fig. 1(b). A short-circuit branch is set at $x = x_f'$ along the line as the guessed fault location. The expression for the short-circuit current in the frequency domain is

$$I_f(x_f', \omega) = \frac{(1+\rho_0)^2 e^{-\gamma(x_f'-x_f)}}{Z_0(1+\rho_0 e^{-2\gamma x_f'})(1+\rho_0 e^{2\gamma x_f})} U_f^*(\omega). \quad (3)$$

Parseval's theorem is used to calculate the short-circuit current energy, which reaches the maximum when $x_f' = x_f$. It is worth noting that the time domain signal can be directly used to calculate the energy in practical application:

$$E = E(x_f') = \sum_{j=1}^{n} I_f^2(x_f', j)\Delta t, n = \frac{T}{\Delta t}, \quad (4)$$

where $T$ is the total duration of the short-circuit current signal, and $\Delta t$ is the simulation time step size. The position of the short-circuit branch corresponding to $E_{\max}$ is the true fault location:

$$x_f = \arg|_{x_f'}(\max(E(x_f'))). \quad (5)$$

Another EMTR fault location method is discussed in [16, 27] and involves calculating the voltage energy along the line. The minimum energy occurs at the mirror point of the fault. However, this method is not suitable for transmission lines with branches because there is no mirror point for the fault.

Consequently, this paper uses the method for calculating short-circuit current energy.

## III. EMTR FAULT LOCATION METHOD FOR POWER LINES WITHOUT BRANCHES

### A. EMTR fault location method using an optimization algorithm

As discussed above, the fault location can be realized by calculating the maximum value of the short-circuit current energy. In the classical method, this is realized by exhausting the position of the short-circuit branch on the transmission line, which requires a huge amount of calculation. The fault location is equivalent to solving the following optimization problem:

$$E_{\max} = \max_{0 < x_f' \leq L} \left( E(x_f') \right), \quad (6)$$

where $L$ is the total length of the line.

The objective function in the optimization problem expressed in (6) has the following characteristics. First, it is difficult to find its analytical expression. Second, the number of extremum points of the function is unknown. The iteration process may fall into a local extremum when we use a traditional optimization algorithm such as the steepest-descent method, and the fault point cannot be accurately located.

This paper uses simulated annealing (SA) because it has better global convergence and does not require the analytical expression of the function or its derivative [26].

The idea of SA originates from the process of solid annealing. 'Annealing' refers to a process of heating an object to a certain temperature and then cooling it. If the temperature is rapidly lowered during solid cooling, the atoms cannot become arranged into a regular crystal structure in time, thereby generating an amorphous crystal. However, if the solid is heated to a higher temperature and then cooled slowly, the atoms can become gradually ordered after gaining energy, thereby eventually forming a crystal structure at room temperature and reaching the most stable state, namely the energy minimum. Kirkpatrick et al. proposed using SA to solve the problem of optimization falling into a local extremum in 1983 [23]. The amorphous state of a solid corresponds to a local extremum, and the crystal state corresponds to the global extremum, which has the lowest energy.

SA mainly includes two parts, the cooling process and the Metropolis criterion [24]. The algorithm parameters include the starting temperature $T_0$, termination temperature $T_t$, cooling coefficient $k$ ($0<k<1$), and termination parameter $N$.

The iterative process starts at a high temperature $T = T_0$. The new solution is generated by randomly disturbing the selected initial solution, which is accepted if it is higher (lower for minimization problems) than the present solution; otherwise, it is accepted with the following probability:

$$P = \begin{cases} 1, & E_{new} < E_{old} \\ e^{-\frac{E_{new}-E_{old}}{T}}, & E_{new} \geq E_{old} \end{cases}. \quad (7)$$

Equation (7) is the Metropolis criterion. $T$ is the current temperature, which drops by $T' = kT$ as the iterative process carries on. Because the starting temperature is high, the algorithm accepts most of the new solutions in the initial stage of the search. As the temperature decreases, the probability of accepting worse solutions gradually decreases, and the searching direction stabilizes accordingly.

In classical SA, temperature drops only after the algorithm converges at the given temperature, which is achieved when no new solution is generated after several iterations, and the final solution at the termination temperature is output as the optimal solution. For the problems and required accuracy in this paper, numerical experiments show it is not necessary to converge at every temperature.

Therefore, we make some adjustments to the algorithm. First, temperature drops once a new solution is accepted. Second, if no new solution is generated after $N$ iterations, it is determined that the algorithm has converged and the current solution is output as the optimal solution.

The SA flow chart is shown in Fig. 2.

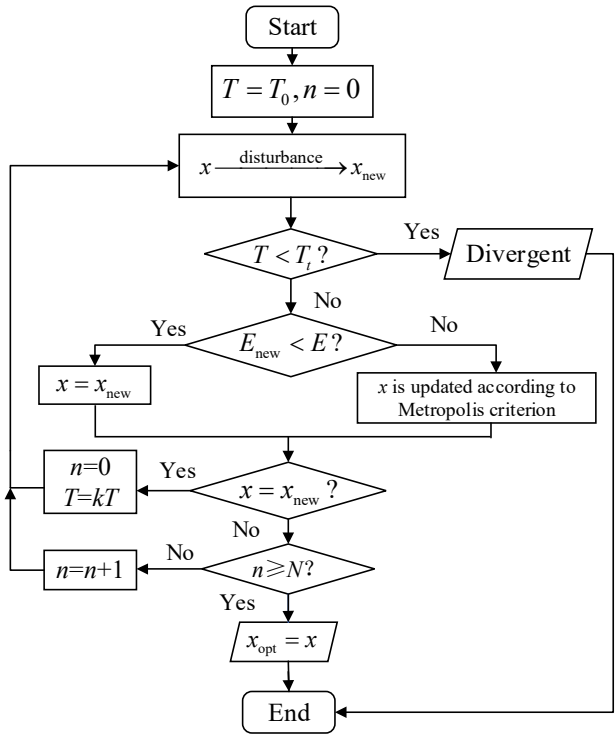

Fig. 2. Flow chart of the SA method.

The Metropolis criterion causes the objective function to accept a worse solution with a certain probability, which retains the possibility of jumping out of a local optimum and continuing to search for the global optimum without remaining trapped in the local extremum.

Therefore, the method performs better at searching for the global optimum. It has been proved that the algorithm has asymptotic convergence [26].

### B. An application example of the optimization method

The short-circuit fault shown in Fig. 1(a) is considered. We assume that the short-circuit fault occurs at the location $x_f = 4$ km. The short-circuit type is set to ideal (short-circuit impedance equals to 0). A 10-kV power-frequency AC voltage is applied to the head end of the line. Both ends are connected with 100-kΩ resistances that serve as the equivalent of power transformers. The circuit parameters are shown in Table I. The simulation step size is set to 0.1 μs.

TABLE I. PARAMETERS OF THE OVERHEAD TRANSMISSION LINE

| Line parameter | Value |
| --- | --- |
| Line length | 10 km |
| Inductance per unit length | 1.60 μH/m |
| Capacitance per unit length | 10.54 pF/m |
| Resistance per unit length | 0.036 mΩ/m |

We use SA to locate the fault. The disturbance is set as a random number within one fifth of the length of the defined function domain in this paper. To accelerate the calculation, the disturbance interval is halved for each iteration if n≥N/2.

When $T_0$ is set to such a small value that the Metropolis criterion almost only accepts better solutions, the SA method degenerates into a random walk optimization (RWO) [29], and therefore may fall into a local optimum. To test this case, we set $k$=0.9 and $N$=10, and set $T_0 = 10^{-9}$ K and $T_0 = 10^{-1}$ K as examples for RWO and SA, respectively.

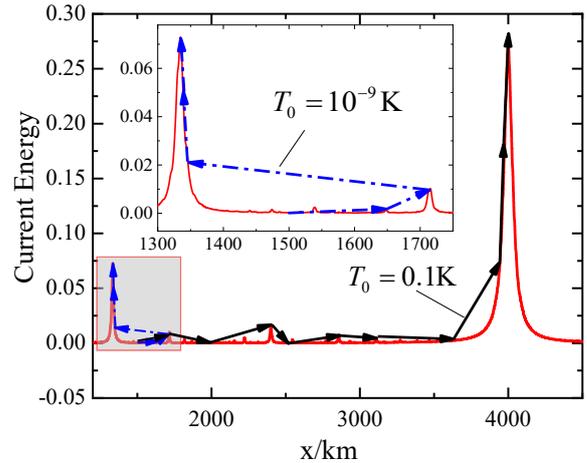

Fig. 3. Iterative process with different starting temperatures: $T_0 = 10^{-9}$ K (RWO) and $T_0 = 0.1$ K.

As shown in Fig. 3, the iterative process starts from the same point $x$=1500 m. SA obtains the global maximum while RWO falls into a local optimum, which demonstrates the effectiveness of SA.

The uncertainty in the search direction and initial value of the algorithm makes the number of iterations random. Therefore, the calculation is repeated 10 times with the accuracy ranging from 5 m to 20 m. The calculation time and number of iterations are shown in Fig. 4.

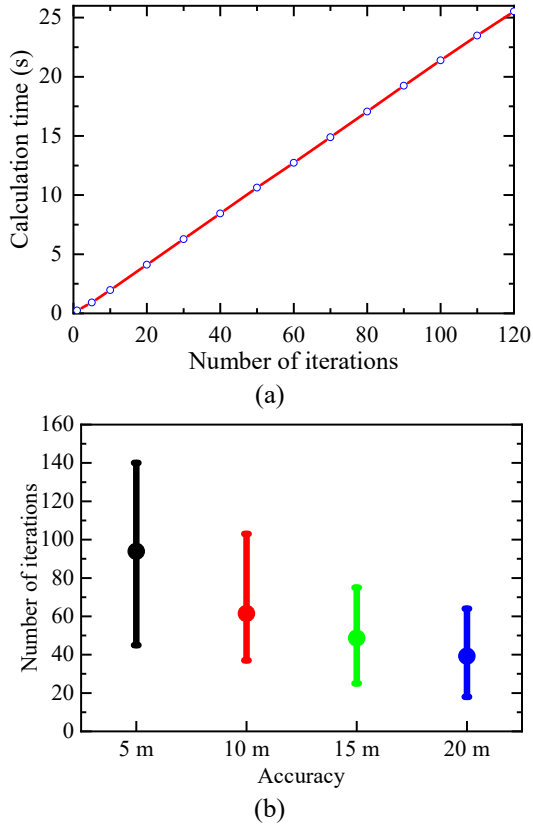

Fig. 4. Calculation time and number of SA iterations with different accuracy requirements. All codes including the transient electromagnetic simulation and SA algorithm are implemented using MATLAB.

The calculation times shown in Fig. 4(a) are proportional to the number of iterations, which indicates that the efficiency can be represented by the number of iterations. When the locating accuracy is set to 5 m, the average number of SA iterations is 93 as shown in Fig. 4(b), while 2000 calculations are required for the exhaustive method. The calculation speed of classical EMTR fault location is more than 10 times faster using SA. Therefore, the optimization algorithm greatly improves the locating efficiency.

## IV. EMTR FAULT LOCATION METHOD FOR POWER LINES WITH BRANCHES

A distribution network may contain many power lines with branches. It is difficult to use a single variable to traverse all possible paths in the network. Therefore, the fault location problem cannot be transformed into a univariate optimization problem as in (6).

A complex network of power lines with branches can be regarded as a connected graph. In this section, the network is divided into several one-dimensional simple lines on the basis of graph theory. Therefore, the fault can be located by solving the optimization problem on each line.

### A. Fundamentals of Graph Theory

**Definition I (graph).** The tuple $G = (V, E)$ is called a graph, where $V$ is the set of nodes and $E$ represents the edges between nodes.

**Definition II (path).** $P = \{v_{i1}, e_{i1}, v_{i2}, e_{i2}, \cdots e_{iq}, v_{iq+1}\}$ is called a path in $G$ if $v_{ik}$ and $v_{ik+1}$ are two end nodes of the edge $e_{ik}$ in $G = (V, E)$.

**Definition III (connected graph).** $G$ is called a connected graph if there is a path between any two nodes in $G$.

In general power systems, isolated nodes have no connection with other parts of the network without considering electromagnetic radiation or coupling, which is not necessary for analysis under normal circumstances. Therefore, all graphs studied in this paper are connected graphs.

**Definition IV (node degree).** The number of edges associated with a node $v$ of the graph $G = (V, E)$ is called the degree of the node, which is expressed as $d(v)$.

A node with an odd (even) degree is called an odd (even) node.

**Definition V (bridge).** An edge $e_i$ is called a bridge if the graph is not connected anymore after $e_i$ is deleted in $G = (V, E)$.

The following theorem holds under Definitions I–V:

**Theorem I.** The number of odd nodes in the graph $G = (V, E)$ must be even.

*Proof.* Suppose that there are $n$ nodes and $m$ edges in an arbitrary graph $G = (V, E)$. The contribution of $m$ edges to all node degrees is $2m$ because the contribution of each edge to the degree of nodes $u$ and $v$ is 1.

We suppose that $V_e$ is the even node set of graph $G$, and $V_o$ is the odd node set of graph $G$. Then we have

$$\sum_{v \in V_e} d(v) + \sum_{v \in V_o} d(v) = 2m . \qquad (8)$$

Therefore, $\sum_{v \in V_o} d(v)$ is an even number, which indicates $G$ contains an even number of odd nodes. ∎

**Theorem II.** Deleting all edges in a path reduces the number of odd nodes in $G = (V, E)$ by a maximum of 2.

*Proof.* A node in a path is either internal or terminal.

For internal nodes, 2 edges are deleted, which has no effect on the parity of the degree.

For the terminal nodes of the path, only one connected edge is deleted, which changes the parity of the degree. There are three possible parity combinations:

(1) The degrees of both terminal nodes are odd.

The deletion changes the degrees of the 2 nodes from odd to even, which reduces the number of odd nodes by 2.

(2) One terminal node is odd and the other is even.

The deletion changes the odd degree into even and the even into odd. Consequently, the number of odd nodes keeps unchanged.

(3) The degrees of both nodes are even.

The deletion changes the degrees of the 2 nodes from even to odd, which increases the number of odd nodes by 2.

Therefore, the deletion reduces the number of odd nodes by a maximum of 2. ∎

**Theorem III.** A graph with $2k$ ($k \in \mathbb{N}^*$) odd nodes cannot be decomposed into less than $k$ paths.

*Proof.* According to Theorem II, deleting edges in a path can reduce the number of odd nodes by a maximum of 2. Therefore, the number of odd nodes can be reduced by no more than $2(k–1)$ with less than $k$ paths. This indicates there are still odd nodes left in the graph, which means the graph is not completely decomposed. ∎

**Theorem IV.** A graph with $2k$ ($k \in \mathbb{N}^*$) odd nodes can be decomposed into $k$ paths.

*Proof.* (Mathematical Induction)

(i) For *k*=1, the graph is a semi-Euler graph; therefore, the problem degenerates into finding the Euler path [25], which was solved already.

(ii) We assume the theorem holds for *k* and consider the case *k*+1.

A path that both starts and ends with an odd node is selected, and the edges in it are deleted. We know from Theorem II that the number of odd nodes decreases by 2 [from 2(*k*+1) to 2*k*]. According to the inductive hypothesis, the remaining part of the graph can be decomposed into *k* paths. Therefore, the entire graph can be decomposed into *k*+1 paths. ∎

According to Theorems III and IV, a graph with $2k$ odd nodes can be decomposed into $k$ paths, and $k$ is optimal.

### B. Method of decomposing complex connected graphs

We now present a practical decomposing method.

The adjacency matrix **A** of the graph is defined as an *n*\**n* matrix whose elements are defined as

$$a_{ij} = \begin{cases} 1, & e_{ij} = (v_i, v_j) \in E \\ 0, & \text{else} \end{cases} . \quad (9)$$

It is easy to conclude that **A** is symmetric, and the degree of node *i* equals $\sum_{j=1}^{n} a_{ij}$.

A path both starting and ending with an odd node is selected randomly from the graph. Then the edges in the path are deleted. The elements $a_{ij}$ and $a_{ji}$ in the adjacency matrix of the graph are zeroed to target each edge in the path accordingly. The path is regarded as the first path of decomposition.

We perform the above operation *k*–1 times and obtain *k*–1 decomposition paths. Meanwhile, the number of odd nodes in the graph $G = (V, E)$ is reduced by 2(*k*–1) according to Theorem II. Hence, there will be only 2 odd nodes left in the graph, making the remaining graph a semi-Euler graph, which can be traversed using one Euler path.

An efficient algorithm to obtain the Euler path was proposed by Fleury in 1921[28], which we refer to as Theorem V.

**Theorem V (Fleury Algorithm).** Given a semi-Euler graph $G = (V, E)$, the algorithm for finding an Euler path in *G* is as follows.

(1) An odd node $v_0$ is selected such that $P = \{v_0\}$.
(2) Assuming the path $P = \{v_0, e_1, v_1, e_2, \cdots e_i, v_i\}$ is determined, $e_{i+1}$ is chosen according to the following principles:
   (a) $e_{i+1}$ is connected with $v_i$;
   (b) $e_{i+1}$ should not be one of the bridges in $G = (V, E)$ unless there is no other choice.
(3) The algorithm ends if (2) cannot proceed anymore.

Theorem IV establishes a specific way of implementing the *k*-th decomposition path. It is worth noting that the decomposition is not unique because the selection of the initial nodes and the path is not unique.

We give two examples to illustrate the correctness of the decomposing method. The Chinese character "Chuan" is shown in Fig. 5(a). There are only two odd nodes, node 1 and node 14, so the graph is a semi-Euler graph, which can be decomposed into a single path. A possible way given by the Fleury algorithm is illustrated in Fig. 5(b).

Fig. 6(a) shows an abstracted graph derived from the Konigsberg Seven Bridges Problem. There are 4 odd nodes, namely 1, 4, 7, and 8 in the graph. A decomposition is shown in Fig. 6(b). The graph is decomposed into two paths accordingly. In spite of the complexity of the graphs in Figs. 3 and 4, they can still be decomposed in an optimal way, which proves the decomposing method is correct.

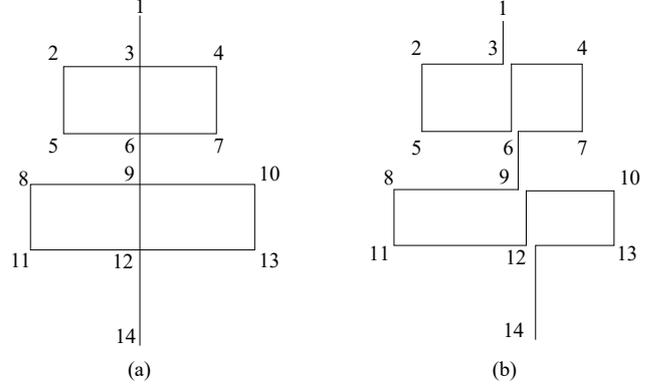

Fig. 5. Connected graph of the Chinese character 'chuan' and its decomposition into a single path.

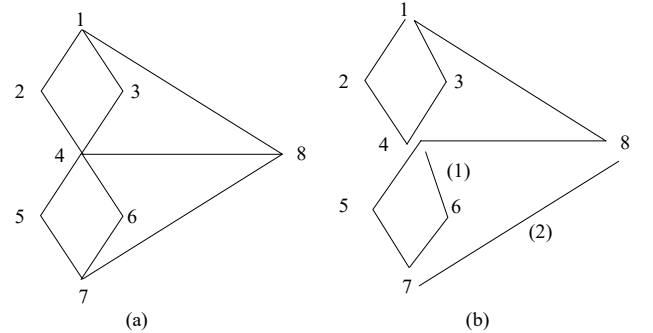

Fig. 6. Connected graph of the Konigsberg Seven Bridges and its decomposition into 2 paths.

### C. EMTR fault location method for power lines with branches

The topology of transmission lines with branches can be abstracted into a connected graph with $2k$ odd nodes. The decomposing method described above is used. The *k* paths obtained from the decomposition are actually *k* one-dimensional power lines. Only one variable is needed to traverse all the positions on every single line. Therefore, (6) can be applied to each line *i* using the optimization algorithm

$$E_{i,\max} = \max_{0 < x_f' \leq L_i} \left( E_i(x_f') \right), i = 1, 2, \cdots, k \ , \quad (10)$$

where $L_i$ is the length of the one-dimensional transmission line *i*. The maximum value of all $E_{i,\max}$'s in (10) corresponds to the location $x_f$ of the real fault point:

$$x_f = \arg|_{x_f'} \left( \max_{1 \leq i \leq k} (E_{i,\max}) \right) . \quad (11)$$

Table II describes the process of EMTR fault location for transmission lines with branches.

TABLE. II. PROCESS OF EMTR FAULT LOCATION FOR POWER LINES WITH BRANCHES

| |
|---|
| **Input**: network parameters; network topology and voltage signal $U_0$ generated by fault |

1. $U_0$ is reversed in time domain.
2. The connected graph with $2k$ odd degree nodes abstracted from the network topology is decomposed into $k$ one-dimensional lines.
3. For each one-dimensional line ($1 \leq i \leq k$), set a short-circuit branch at $x_f'$ as the guessed fault position.
4. The signal is re-injected into the network. Calculate the short-circuit current energy $E_i(x_f')$ for $1 \leq i \leq k$.
5. Solve the optimization problem $E_{i,\max} = \max\limits_{0 < x_f' \leq L_i}(E_i(x_f'))$ for $1 \leq i \leq k$ using SA.
6. Find $E_{\max} = \max\limits_{1 \leq i \leq k}(E_{i,\max})$.

**Output**: the predicted fault location is $x_f = \arg|_{x_f'}(E_{\max})$.

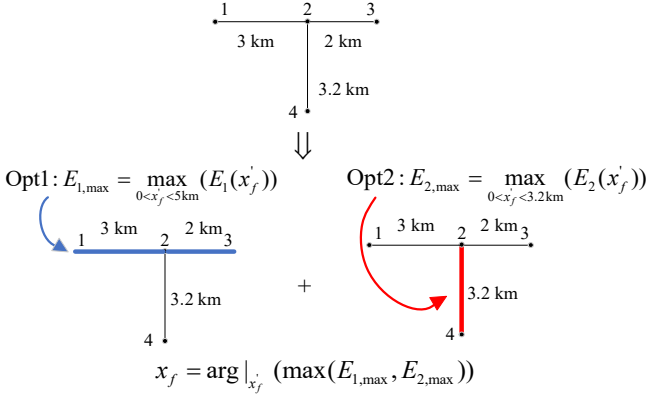

Fig. 7. Optimization problems and the real fault position of a T-type network.

We give an example to illustrate the EMTR fault location method. The abstract connected graph of a T-type network is shown in Fig. 7. The transient voltage generated by the fault is collected at node 1 and re-injected into the original network after time reversal. All nodes in Fig. 7 are odd, so the graph shall be decomposed into at least two paths. One way to do this is: ① 1–2–3, ② 2–4. Therefore, two optimization problems need to be solved as shown in Fig. 7. We emphasize the transient simulation is performed in the original power network.

## V. APPLICATION EXAMPLES

We consider a distribution network of power lines containing 11 nodes totally, as shown in Fig. 8.

Part of the network above node 4 is cable-connected, which is bold in Fig. 8. The overhead line and cable parameters are shown in Tables I and III.

A 10-kV power-frequency AC voltage is applied to node 1. The terminals of the network are connected to power transformers, which are equivalent to a large impedance. We adopt 100 kΩ in the example [3]. A short-circuit fault is set along the line. The transient voltage signal generated by the fault is obtained at node 1, and then injected into the system after time reversal. The simulation step is 0.1 μs.

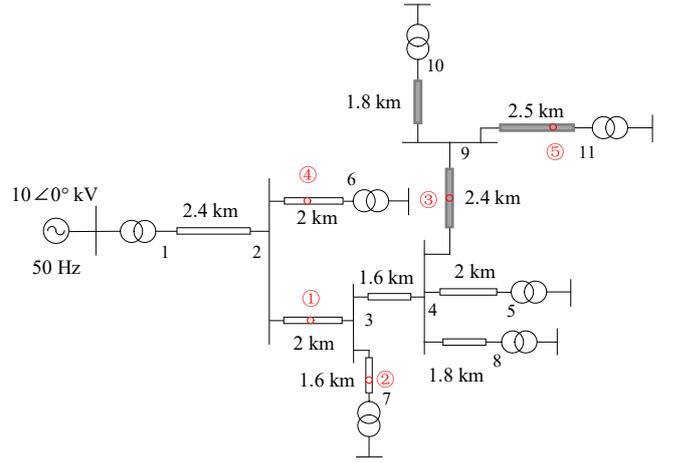

Fig. 8. Power distribution network model (containing 11 nodes) with different fault positions.

TABLE. III. CABLE PARAMETERS

| Line parameter | Value |
| --- | --- |
| Inductance per unit length | 0.583 μH/m |
| Capacitance per unit length | 201 pF/m |
| Resistance per unit length | 0.008 Ω/m |

There are 10 odd nodes in the graph, which indicates the network should be decomposed into five one-dimensional lines. One way to realize this is:

① 1–2–3–4–5, ② 2–6, ③ 3–7, ④ 8–4–9–10, ⑤ 9–11.

Therefore, the guessed short-circuit branch is set on five one-dimensional transmission lines. Then, SA is used to solve the corresponding optimization problem with the short-circuit current energy as the objective function. We set $N$=10 in Fig. 2. The starting temperature is 1 K and the cooling coefficient is 0.8.

Different fault conditions are adopted to analyze the effectiveness and efficiency of EMTR fault location compared with the classical exhaustive search.

### A. Fault position

As shown in Fig. 8, different fault positions are chosen: ① 0.8 km from node 2; ② 1 km from node 3; ③ 1.2 km from node 4; ④ 0.6 km from node 2; ⑤ 2 km from node 9. The fault angle is set to 90°, and the fault impedance is 1 Ω.

The impedance of the short-circuit branch that is the guessed fault position in the EMTR method is set to 20 Ω.

TABLE. IV. CALCULATION RESULT FOR MAXIMUM SHORT-CIRCUIT CURRENT ENERGY AT FAULT ③ (ACCURACY: 10 M)

| Line No. | Maximum point/m | Maximum value/A²·μs |
| --- | --- | --- |
| 1 | 5720 | 2.6781 |
| 2 | 850 | 0.0782 |
| 3 | 500 | 0.0412 |
| 4 | 3000 | 3.1224 |
| 5 | 1870 | 1.0917 |

Table IV shows the calculation results for the short-circuit current energy, which reaches the maximum at 3 km on line 4 (the real fault point). The result shows that the fault can be accurately located.

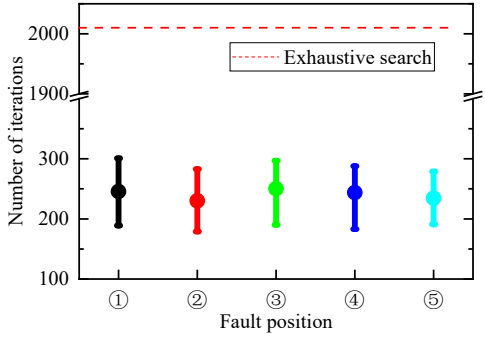

Fig. 9. SA iteration number with different fault positions ($T_0 = 1$ K).

As shown in Fig. 9, the efficiency (represented by the number of iterations) is not sensitive to the fault position. The number of iterations required is about 250 for every fault position, compared with 2010 iterations in the traditional exhaustion method. This shows that the optimization algorithm improves the calculation efficiency by nearly one order of magnitude.

### B. Fault angle and fault impedance

The fault angle and fault impedance affect the amplitude of the transient electromagnetic signal. A small fault angle or high fault impedance produces a small amplitude, which results in less short-circuit current energy.

In this case, the angle is set to 30, 60, and 90 degrees respectively. An accuracy of 10 m can be achieved at these fault angles using EMTR. Fault position ① is chosen, and the number of iterations is shown in Fig. 10 under the same SA parameters.

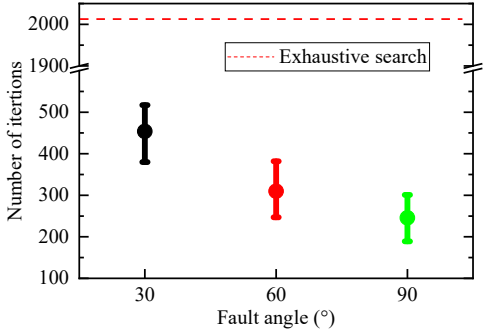

Fig. 10. SA iteration number with different fault angles ($T_0 = 1$ K).

Fault position ① is chosen, and the fault angle is set to 90°. The fault impedance ranges from 0 to 50 Ω. The SA parameters remain the same.

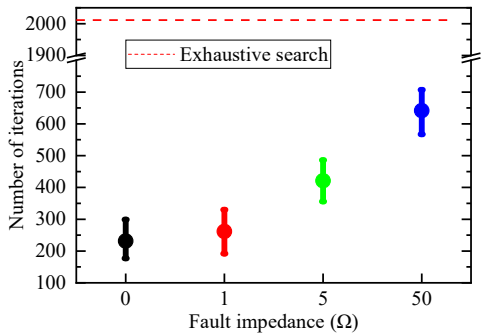

Fig. 11. SA iteration number with different fault impedances ($T_0 = 1$ K).

Fig. 11 shows the number of iterations under different fault impedances. We see that the calculation slows down as the fault impedance increases.

### C. Discussion

The number of iterations increases with decreasing fault angle and increasing fault impedance. According to (7), when the amplitude of the short-circuit current energy decreases, there is a higher probability of accepting worse solutions in SA, which increases the time to achieve convergence. Therefore, the selected starting temperature is related to the amplitude of the objective function, and has a direct impact on the calculation speed.

The transient amplitude is small for a fault of 50 Ω, and therefore requires a lower temperature to maintain the calculation speed. Fig. 12 shows the number of iterations when the starting temperature is changed from 0.01 K to 1 K.

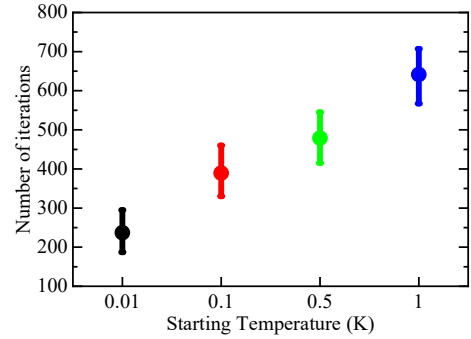

Fig. 12. SA iteration number with different starting temperatures (the fault impedance is 50 Ω).

In this case, the calculation speed is nearly 10 times faster when the starting temperature is 0.01 K. An adaptive selection of the starting temperature for different problems is necessary for accelerating the method and needs further study.

## VI. CONCLUSION

We have proposed an optimization-accelerated EMTR-based fault location method for complex power lines with branches.

The EMTR fault location method solves an optimization problem theoretically. Rather than use an exhaustive search, we use SA to solve the optimization problem. This improves the efficiency by a factor of up to 10 on average.

A network of power lines with branches can be regarded as a connected graph. The difficulty in directly using the SA optimization algorithm for such a case is that it is difficult to use a single variable to traverse all possible fault locations. Therefore, we proposed a method to decompose the network into the fewest simple one-dimensional lines. We proved that a connected graph must contain an even number ($2k$) of odd nodes and can be decomposed into $k$ paths, where $k$ is optimal. In addition, a practical method was proposed in detail to automatically execute the decomposition. Hence, our optimization algorithm reduces to solving several one-dimensional optimization problems.

Numerical experiments were carried out in a distribution network model to illustrate the effectiveness of the method. Results under different conditions show the method works reliably.

Finally, this optimization-accelerated method is also applicable to other EMTR-based fault location methods that seek an optimum, such as the high-impedance fault location method [30].